# Building Resilience to Misinformation: A Cross-National Development of the Digital Media and Information Literacy Scale (DMILS)

Sijia Qian (Corresponding Author)[1], Cuihua Shen[2], Huiyi Wang[3], Hichang Cho[4]

[1] Assistant Professor, School of Data Science, University of North Carolina at Charlotte; Email: sqian@charlotte.edu

[2] Professor, Department of Communication, University of California, Davis; Email: shencuihua@gmail.com

[3] Assistant Professor, Research Center of Journalism and Social Development, School of Journalism and Communication, Renmin University of China; Email: wanghuiyi314@ruc.edu.cn

[4] Associate Professor, Department of Communications and New Media, National University of Singapore; Email: hichang_cho@nus.edu.sg

Abstract

Amid growing concern about information quality and credibility in digital media environments, researchers and educators still lack a concise, comprehensive yet psychometrically sound instrument for tracking the competencies that help people navigate this landscape. This article develops the Digital Media and Information Literacy Scale (DMILS)—a robust and multidimensional measure that distinguishes domain (digital vs. information/news), competency type (knowledge vs. skill), and is measured through both subjective and objective items. Through two empirical studies with three nationally matched samples in the United States and Singapore (N = 1,498), we developed an 18-item self-report battery and 16-item objective knowledge questions, showing strong structural, convergent, and predictive validity, along with a short form (8 self-report and 8 objective items). By offering a parsimonious yet multidimensional yardstick, DMILS enables rigorous evaluation of media literacy interventions and supplies a common metric for cross-national research, critical for building an information ecosystem resilient to mis- and disinformation.

**Building Resilience to Misinformation: A Cross-National Development of the Digital Media and Information Literacy Scale (DMILS)**

In an increasingly complex media landscape, where misinformation spreads rapidly, media literacy has emerged as an essential competency for democratic deliberation and individual well-being (Mihailidis & Thevenin, 2013; Potter, 2019). Although the urgency of media literacy has been widely acknowledged by educators, policymakers, and scholars alike, the field remains conceptually fragmented and empirically inconsistent. Multiple literacies—digital literacy, information literacy, news literacy, algorithm literacy, and AI literacy—have evolved to address specific features of the digital ecosystem, yet they often operate in parallel rather than in concert (Martens, 2010). Similarly, while a growing body of research highlights the promise of media literacy as a tool to reduce misinformation susceptibility (e.g., Guess et al., 2020; Qian et al., 2023), the empirical evaluation of such efforts has been hampered by the absence of a unified, multidimensional measurement framework. Existing scales are often fragmented, outdated, or narrowly focused, failing to capture the full spectrum of relevant competencies or to adapt to rapidly changing technological affordances (e.g., Hargittai & Hsieh, 2012). Although different components of media and information literacy (e.g., news literacy, digital literacy) may play distinct roles depending on the type of content or task, a more unified and parsimonious measure would address these limitations by improving conceptual clarity, reducing implementation costs, and enhancing comparability across studies, while preserving meaningful distinctions among subdimensions.

This paper addresses this gap by introducing a multidimensional framework and a newly developed scale for measuring Digital Media and Information Literacy (DMIL). The framework synthesizes prior scholarship across digital, news, and information literacy, distinguishing between types of media (digital vs. news/information) and types of competence (knowledge vs. skill), each measured through both subjective (self-reported) and objective (performance-based) indicators. Drawing on this framework, we developed and validated the Digital Media and Information Literacy Scale (DMILS) through two studies using three national samples from the United States and Singapore (N = 1,498). These studies followed rigorous scale development and validation procedures, including exploratory and confirmatory factor analyses, Rasch modeling, and tests of convergent and predictive validity.

Taken together, the DMILS offers researchers, educators, and policymakers a psychometrically robust and globally applicable tool for assessing digital media and information literacy. It enables rigorous evaluation of existing media literacy interventions, surfaces specific competence gaps that curricula can target, and provides a common yardstick for comparing programs across cultural contexts.

## The Need for an Integrated Literacy Framework

The rise of mis- and disinformation, particularly in the form of AI-generated deepfakes and synthetic content, has made media- and information-related literacies more urgent than ever (Bulger & Davison, 2018). False political stories spread significantly faster than true ones on social media platforms like X/Twitter (Vosoughi et al., 2018). While detection technologies offer some support, they are insufficient without users' understanding of how such tools work and practical strategies for content verification (Wardle & Derakhshan, 2017). Media-literate individuals are more likely to engage in fact-checking and verification behaviors, demonstrating the ability to actively seek out and assess accurate information in online environments (Khan & Idris, 2019). Prior work shows that objective verification skills strongly predict discernment of true and false news (Jones-Jang et al., 2021) and that digital media literacy is the only meaningful defense at the individual level against fake and manipulated news images (Shen et al., 2019). Cross-national evidence with adolescents indicates that specific digital-interaction skills lower self-reported misinformation exposure (Vissenberg et al., 2023).

Despite this growing consensus that media and information literacy plays a critical role in mitigating misinformation, the field has yet to converge on a shared conceptual and measurement framework. Researchers often use different terms to describe similar constructs. Constructs such as awareness and cognitive processing, though labeled differently, often reflect overlapping aspects of media-related knowledge (Oeldorf-Hirsch & Neubaum, 2025). Despite the growing number of interventions promoting literacy, many programs invoke the concept without rigorously measuring it (e.g., Kahne & Bowyer, 2017).

Second, many existing measures capture only one dimension of media and information literacy, rather than its full breadth. For instance, some measures focus exclusively on skills while neglecting knowledge, or vice versa (e.g., Livingstone & Helsper, 2010). Skills and behaviors related to literacy (e.g., verification strategies) are frequently assessed separately from knowledge without a scale that integrates both dimensions (Vraga & Tully, 2021; Vraga et al., 2021). The problem with this piecemeal approach is that a comprehensive measurement of media and information literacy would have to incorporate numerous individual dimensions, resulting in an extremely lengthy and convoluted scale with substantial redundancy and item overlap. While a few relatively comprehensive, multidimensional scales are available (e.g., New Media Literacy scale, Koc & Barut, 2016), they often focus on specific contexts and tend to include an extensive number of measurement items. For practical purposes, we need a parsimonious yet multidimensional scale.

Third, many prominent literacy scales were developed over a decade ago (e.g., Hargittai & Hsieh, 2012), limiting their relevance in rapidly evolving digital environments involving advanced algorithms and AI-driven tools. Notably, existing scales have predominantly focused on Western contexts (Livingstone & Helsper, 2010; Ng, 2012), raising concerns about cultural validity and relevance in non-Western contexts. A scale developed with U.S. adults in mind, for instance, might assume familiarity with certain media outlets, value orientations (like a press-freedom perspective) that do not fully translate to other societies.

Lastly, many existing measures rely heavily on self-reported, subjective assessments. Research consistently shows that perceived and actual knowledge are distinct constructs (Radecki & Jaccard, 1995), and confidence in one's abilities does not necessarily reflect true competence. Incorporating both types of measures enables researchers to identify misalignments and generate a more accurate understanding of literacy levels.

## Conceptualizing Digital Media and Information Literacy (DMIL)

We propose a new multidimensional framework for Digital Media and Information Literacy (DMIL) that captures how individuals understand and engage with digital technologies and information content in contemporary media environments. DMIL is structured along two core axes: (1) *domain*—digital media versus information/news—and (2) *competency type*—knowledge versus skill (see Table 1). This structure reflects a process-based view of media literacy. Media literacy theory conceptualizes literacy as a structured process rather than a single, undifferentiated capacity. Foundational accounts distinguish among competencies related to how individuals encounter, interpret, and evaluate content (Potter, 2019). These competencies are analytically separable because they perform distinct functions: knowledge components organize individuals' understanding of how media and information systems operate, skill components determine whether individuals can act within those systems. From this theoretical view, no single dimension can fully stand in for the others, as each corresponds to a different stage in the literacy process and addresses a different potential outcome. Next, we explicate the domain and competency-type distinctions that organize DMIL, and describe how these components are operationalized within the framework.

### ***Domains: Digital Media and Information Literacy***

The digital media domain of DMIL is grounded in a broad body of conceptually-overlapping scholarship in digital literacy (Cordell, 2013), ICT (information and communication technology) literacy (Olsson et al., 2019), and new media literacy (Ng, 2012). Collectively, these concepts focus on individuals' ability to access, evaluate, generate, and use content through digital technologies, combining technical proficiency with cognitive and evaluative competencies such as assessing credibility and accuracy. New media literacy further extends these foundations by incorporating critical and participatory dimensions (Chen et al., 2011).

More recent scholarship has extended this line of literature to the socio-technical affordances that define today's media ecology. Social media literacy foregrounds users' understanding of platform architectures and their ability to manage self-presentation, evaluate user-generated content, and fact-check influencer claims (Cho et al., 2024), algorithm literacy emphasizes the ability to understand and critically engage with algorithmic systems (Dogruel et al., 2022; Oeldorf-Hirsch & Neubaum, 2025), and the rise of generative AI has sparked calls for AI literacy (Wang et al., 2023). Despite their issue-specific foci, these literacies still preserve the same spine as digital media literacy.

The information and news domain of DMIL draws from longstanding traditions in media literacy, information literacy, and news media literacy that focus on journalistic content and verification practices. Media literacy theory argues that understanding how messages are constructed reduces susceptibility to distorted portrayals of reality (Christ & Potter, 1998). Extending this logic, news media literacy is defined as the knowledge and skills required to navigate news environments critically (Ashley et al. 2017), including awareness of production processes, economic incentives, and interpretive biases (Maksl et al., 2015).

DMIL integrates both digital media and information domains because today's information challenges are not neatly confined to a single domain. A holistic framework like DMIL reflects the convergent nature of these competencies and provides a more comprehensive foundation for interventions, assessments, and education.

***Competency Types: Knowledge and Skills***

The competency type dimension distinguishes between knowledge and skills**,** a distinction that is as conceptually well supported by prior research. Potter's (2019) framework highlights knowledge structures alongside analytical and evaluative skills as core building blocks of media literacy. Similarly, Ng (2012) explicitly separates technical knowledge (e.g., understanding how browsers route traffic) from technical skills (e.g., configuring privacy settings). Extending this line of work, we do not treat knowledge and skills as abstract or general capacities. Instead, we define them in relation to specific media and information domains. Together, these competencies support key outcomes such as discernment (the accuracy with which individuals distinguish true from false content) and information retrieval accuracy (the ability to locate and extract accurate information online), which serve as the primary dependent variables predicted by DMIL in the present study (see Measurement Model in Figure 1).

***Operationalization of DMIL***

Within the conceptual framework, **digital knowledge** is measured using both self-reported familiarity with technology-related terminology (e.g., PDF, hashtag, algorithm) and objective items testing their understanding of these terms. Such knowledge reflects a user's practical grasp of the tools, structures, and conventions that underpin everyday online information activities. Importantly, Hargittai and Hsieh (2012) demonstrate empirically that respondents who report greater familiarity with these terms also perform significantly better on objective information-seeking tasks, thereby shaping the conditions under which discernment can occur. We used the terms from this validated instrument and augmented them with contemporary items, such as "algorithms," to reflect the evolving digital landscape.

**Digital skill** is measured through self-reported ability to operate, navigate, and create content using digital tools and technologies. These operational competencies directly support discernment by enabling users not only to access relevant information, but also to compare sources, verify claims, and recognize inconsistencies or contextual cues that signal misleading or low-quality content (Guess & Munger, 2023; Vraga & Tully, 2021).

**Information knowledge** is measured through both self-reported understanding of how news and information are produced, structured, and shared, and objective items (multiple-choice questions) that assess respondents' knowledge of these processes. Such understanding enables users to contextualize claims and assess source credibility (Vraga et al., 2021).

Finally, **information skill** is measured through self-reported ability to critically evaluate news and information, verify content, and engage with media through informed judgment or social interaction. It represents the most proximal competency for discernment, as it directly involves the critical evaluation of evidence. Consistent with prior research, evaluative and analytic skills have been shown to be stronger predictors of misinformation detection than general knowledge alone (Amazeen & Bucy, 2019).

## Overview of Studies

Our goal was to develop a parsimonious, comprehensive, and up-to-date Digital Media and Information Literacy Scale (DMILS) that reflects contemporary and emerging digital information development. The scale includes both subjective (S-DMILS) and objective (O-DMILS) components, building on our multidimensional framework. The scale was tested and validated in both the United States and Singapore to ensure its international relevance. These two countries share several key socio-economic characteristics—including high levels of economic development, advanced ICT infrastructure, and the widespread use of English as an official language—which make cross-national comparisons particularly meaningful by reducing potential confounding factors. These two contexts also provide an instructive contrast. The United States features a highly decentralized media environment characterized by intense political polarization and declining trust in mainstream news outlets and government institutions (Barrett et al., 2021). In contrast, Singapore's media environment is more centralized and regulated, and surveys generally report comparatively higher levels of trust in mainstream media and official information sources. For example, the Reuters Digital News Report (Newman et al., 2025) shows that overall trust in news is substantially higher in Singapore (45%) than in the United States (30%). This institutional structure is accompanied by government-led initiatives promoting discernment and "cyber wellness" as part of national policy (Ministry of Education Singapore, 2025). Validating the scale in two contexts is important, as issues like misinformation and the adoption of AI are global in scope, and there is increasing demand for cross-national evidence on media literacy intervention outcomes (Hoes et al., 2024). We followed an iterative scale development process and used a two-study process to refine and validate the measure. The studies were pre-registered on the Open Science Framework (OSF, https://osf.io/eysgq/?view_only=abc08d2817c74936b22686f045199f05).

## Study 1

We devised a three-step process for scale development: (1) item collection, (2) item review and reduction, and (3) psychometric analysis, following established guidelines for scale development (Boateng et al., 2018; Carpenter, 2018; Clark & Watson, 1995).

### Methods

#### *Item Collection, Review, and Reduction*

A comprehensive literature search was conducted to identify existing quantitative scales related to digital media and information literacy, using a journal-based approach across 26 journals and conference proceedings in communication and computing fields. Using a broad set of literacy-related search terms, the initial search yielded 476 articles, of which 74 empirical studies with relevant scale measures were retained after screening based on predefined inclusion criteria. See Supplementary Materials (SM) Section 3 for more details.

The authors evaluated the measures collected from the previous step and categorized them into the proposed multidimensional framework, including six cells of competencies (e.g., digital knowledge). We closely inspected each item to determine its domain, type of competency, and mode of measurement, ensuring alignment with the conceptualization of each dimension. When making selection decisions, we considered popularity, recency, and target audience. We deleted items that were repetitive and overly specific to a single-country context. In addition, we revised some items to enhance their relevancy to emerging technologies. This process was conducted through multiple rounds of review and discussion among the authors until consensus was reached.

To address the limited availability of objective digital knowledge measures, we constructed a pool of contemporary digital media terms and generated corresponding true/false items using ChatGPT-4o, supplemented by items adapted from prior validated scales (see details in SM Section 3).

The final pool of items includes 56 subjective items and 30 objective items assessing digital media and information literacy (see description of specific items in SM Section 4).

### *Participants*

To evaluate the reduced item pool, we conducted an online survey with participants recruited from Prolific (N = 499), a crowdsourcing platform commonly used in behavioral research. Participants received $4 for participating in the survey.

To be eligible, participants had to be at least 18 years old and reside in the United States. Sampling relied on quotas for sex, age, ethnicity, and political affiliation based on simplified US census data. In terms of the racial composition of the sample, 315 participants were Non-Hispanic White (63.1%), 68 were African American (13.6%), 47 were Hispanic/Latino (9.4%), 44 were Asian (8.8%), 1 was Hawaiian or other Pacific islander (0.2%), 6 were American Indian or Alaskan Native (1.2%), and 7 were Others (1.4%). Out of the 499 participants, 48.5% identified as Women, and the median age group was “45-54”.

## Results

### *Subjective Measures*

To evaluate the suitability of the subjective measures (56 items) for factor analysis, we calculated the Kaiser-Meyer-Olkin (KMO) measure of sampling adequacy and Bartlett’s test of sphericity with the "EFAtools" R package, following scale development guidelines recommended by Carpenter (2018). The overall KMO value was 0.94 and Bartlett's test of sphericity was significant ($\chi^2(1540) = 14{,}985.02$, $p < .001$), indicating that the items were sufficiently suited for factor analysis.

We then performed parallel analysis with the "psych" R package to determine the number of factors. This analysis involved selecting a parsimonious factor structure where each factor had eigenvalues exceeding the 95th percentile of corresponding eigenvalues from 500 simulated random datasets. The parallel analysis on the 56 items suggested a seven-factor solution (eigenvalues: F1 = 15.45, F2 = 3.37, F3 = 2.65, F4 = 2.31, F5 = 0.87, F6 = 0.73, F7 = 0.68; they are all larger than the simulated random value). Four factors explained the most variance and have eigenvalues larger than 1, which is in line with our theoretical model. An exploratory factor analysis (EFA) using unweighted least squares (ULS) estimation showed that the other three factors had low factor loadings (<.30). We therefore continued with the four-factor structure.

Following this, we conducted an EFA using principal axis factoring with oblique rotation, given the theoretical expectation of a positive correlation between these factors. We applied the following exclusion criteria: (1) factor loadings below .40 (Clark & Watson, 1995), (2) cross-loadings above .30 (Boateng et al., 2018), (3) communalities below 0.4 (Carpenter, 2018), and (4) inter-item correlation above 0.7 (Clark & Watson, 1995). Based on the criteria, we conducted rounds of item removal, resulting in a robust and interpretable solution comprising 26 items. Results of the pattern matrix showed support for our four-factor framework (see SM Section 6).

Given the multidimensional nature of the scale and our objective to create a concise yet robust measure, we aimed for a balanced selection of 4–6 items per factor. This approach resulted in a refined scale of 18 items, with factor loadings ranging from 0.58 to 0.85[1]. The final four-factor structure explained 55.5% of the total variance, indicating strong structural validity. The initial internal consistency analyses confirmed excellent reliability (Cronbach's α: Overall subjective DMIL = 0.93; DK= 0.82, DS = 0.85, IK = 0.85, IS = 0.83). See Table 2 for the items and factor loadings.

***Objective Measures***

The difficulty of the 20 items of *objective digital knowledg*e ranged from *P* = 0.49 to *P* = 0.99 (M = .87, SD = .14). *P* indicates the proportion of participants who answered the item correctly; therefore, lower values reflect higher item difficulty. We identified six items with extremely low difficulty ($P \geq 0.98$), suggesting that these items were ineffective in distinguishing between individuals with varying levels of digital knowledge. We opted to revise these items for improved discriminability in Study 2.

The difficulty of the 10 items of *objective information knowledge* ranged from *P* = 0.35 to *P* = 0.98 (M = .69, SD = .20). Similarly, we identified one item with an extremely high correct response rate (*P* = 0.98). To address this, we revised the answer choices in Study 2 to increase the item's difficulty. We also removed one question as the answer only focuses on the U.S. context.

**Discussion**

Study 1 confirmed the proposed four-factor structure of the subjective DMILS, encompassing DK, DS, IK, and IS. The model demonstrated excellent fit, supporting the multidimensional nature of the construct. Beyond validating the structure, we refined the scale by identifying and removing redundant or low-performing items. The subjective component was reduced from 56 to 18 items. For the objective items, we revised or replaced questions based on item difficulty and clarity. We further validated and tested the refined scale in Study 2.

[1] The initial item pool included both positively and negatively worded items; however, only positively worded items met psychometric criteria during EFA and were retained in the 18-item final scale. We acknowledge that this may introduce potential acquiescence bias and discuss this consideration in the SM Section 5.

## Study 2

Study 2 aimed to further validate the newly developed scale by examining three key aspects of its validity. First, we evaluated the construct validity of the scale through confirmatory factor analysis (CFA) and Rasch modeling. Second, we examined the scale's criterion-related (or nomological) validity by testing its theoretical associations with relevant antecedents and outcomes. Third, we assessed the cross-cultural applicability of the scale to determine its robustness and generalizability across different cultural contexts. We also developed a short form of the scale (see Appendix 1). Specifically, we addressed the following research questions: (a) *Is the scale structure valid?* (b) *Does the construct behave as predicted by theory?* and (c) *Is the measure applicable across cultures?*

### Methods

#### *Participants*

Study 2 used two nationally matched samples from the United States (N = 501) and Singapore (N = 498), recruited from Prolific and Bilendi respectively. Both samples were obtained using quota-based, non-probability sampling to approximate national population distributions (U.S.: age, gender, ethnicity, political affiliation; Singapore: age, gender, race/ethnicity). See SM Section 5 for detailed sample breakdowns. For the U.S. sample, among the 501 participants, 63.9% identified as White, 50.7% identified as women, and the median age group was 45–54 years. For the Singapore sample, among the 498 participants, 80.7% identified as Chinese, 9.6% as Malay, 5.4% as Indian, 3.4% as Other ethnicities, and 0.4% preferred not to answer. Additionally, 50.9% identified as women, and the median age group was 45–54 years.

### *Measures*

**Social media use.** Participants reported how frequently they used ten major social media platforms: Facebook, Twitter, Instagram, Snapchat, YouTube, Reddit, TikTok, LinkedIn, Pinterest, and WhatsApp (Oeldorf-Hirsch & Neubaum, 2025), using a 5-point Likert scale (1 = *Never*, 5 = *Always*). We calculated the number of platforms each participant reported using with moderate or higher frequency, defined as a response of 3 (Sometimes) or above (M = 5.24, SD = 2.11). This threshold was chosen to capture at least occasional use (Qian et al., 2023).

**News consumption.** The frequency of news use is measured by asking how many hours per week the participant uses news via various media platforms. Participants chose between zero hours, less than an hour, one to two hours, three to five hours, and six or more hours per week (McWhorter, 2019). These responses were recoded into a five-point scale (M = 2.80, SD = 1.04).

**Discernment.** To evaluate participants' ability to distinguish between true and false content, we measured their discernment of news posts. Participants were asked to rate each post as either real or fake, and scores were calculated based on the number of correct responses out of the 16 posts (M = 10.25, SD = 2.54). The task included 12 text-based and 4 image-based news posts. The text headlines were adapted from the Misinformation Susceptibility Test (MIST) (Maertens et al., 2024) and supplemented with more recent headlines collected from reputable fact-checking and news sources (e.g., Gallup) to improve timeliness and ecological validity. The image-based posts were sourced from Snopes and included examples of authentic images paired with misleading captions. A full list of the headlines is included in SM Section 7.

**Information retrieval accuracy.** As another real-world application of media literacy, participants completed an information retrieval task adapted from Guess and Munger (2023). Participants were presented with three factual questions (e.g., "Who is the Prime Minister of Croatia?") and were explicitly instructed and encouraged to look up the answers using the Internet during the survey. The questions were intentionally obscure to both U.S. and Singaporean participants, thereby minimizing the influence of prior knowledge and ensuring the task reflected actual skills. The final score was calculated as the number of correct answers out of three (M = 2.21, SD = 1).

## Results

### *CFA of Subjective DMILS*

We conducted a CFA using a four-factor structure of the subjective scale[2]. Standardized factor loadings and factor correlations are presented in Figure 2. Although the chi-square statistic was significant, $\chi^2(129) = 371.83$, $p < .001$, this is expected given the large sample size (N = 999). The CFI value of 0.98 and TLI value of 0.97 exceed the conventional threshold of 0.90, indicating excellent incremental fit. The RMSEA was 0.043 (90% CI = 0.038–0.049), indicating close fit. All standardized loadings were significant (p < .01) and above the 0.50 threshold (Hair et al., 2010), ranging from 0.71–0.81 (DK), 0.72–0.87 (DS), 0.75–0.83 (IK), and 0.71–0.83 (IS). AVE values ranged from 0.58 (IS) to 0.64 (DK), exceeding the 0.50 cutoff and the corresponding maximum shared variance (MSV) values (0.24–0.42). Together, these results indicate that the 18-item, four-factor measurement model of subjective DMILS fits the observed data well and demonstrates adequate construct validity.

[2] We assessed whether the sample size was sufficient to recover the proposed measurement model using a Monte Carlo CFA design analysis. Simulated data based on a four-factor, 18-item model were estimated using WLSMV. With N = 1,000, the model converged in all simulations without improper solutions and showed consistently excellent fit (RMSEA = .005, CFI = .999, SRMR = .023), indicating that the sample size is sufficient for reliable estimation of the factor structure.

***Testing a Higher-Order Model of Subjective DMILS***

We also developed a second-order model to examine whether the four first-order factors could be explained by a single higher-order construct (subjective DMILS). This model assumed that the relationships among the four constructs—DK, DS, IK, and IS —stemmed from a broader latent factor and therefore did not allow correlations among them, instead loading them onto subjective DMILS with unidirectional paths. Confirmatory factor analysis indicated a good model fit ($\chi^2(131) = 373.5$, CFI = 0.976, TLI = 0.971, RMSEA = 0.043), with standardized loadings from Subjective DMILS to the four factors at 0.83, 0.78, 0.6, and 0.58, all statistically significant ($p < 0.01$). Additionally, all measurement loadings from the first-order factors to observed items were significant (p < 0.01) and above the recommended threshold of 0.50. A chi-square difference test comparing this model to the correlated first-order model revealed no significant difference ($\chi^2(2) = 1.67$, $p = .43$), suggesting that the more constrained higher-order model fits the data equally well and offers a more parsimonious explanation. Taken together, these results support the interpretation of subjective DMILS as a unidimensional underlying construct.

***Testing Measurement Invariance Across the U.S. and Singapore***

To examine the cross-country applicability, we tested measurement invariance using the 'lavaan' package in R. The configural model, with parameters freely estimated across groups, showed acceptable fit in both groups (U.S. sample: $\chi^2(129) = 326.18$, $p < .001$, CFI = .955, TLI = .947, RMSEA = .055, 90% CI [.048, .063]; Singapore sample: $\chi^2(129) = 244.28$, $p < .001$, CFI = .976, TLI = .972, RMSEA = .042, 90% CI [.034, .050]), indicating equivalent factor structure across countries. For metric invariance, the changes in alternative fit indices supported invariance: ΔCFI = .0032, ΔRMSEA = .0010, and ΔSRMR = .0074. For scalar invariance, intercepts were additionally constrained. Changes in fit indices remained within acceptable limits: ΔCFI = .0079, ΔRMSEA = .0037, and ΔSRMR = .0036. These findings support scalar invariance, indicating that both factor loadings and item intercepts can be considered equivalent across samples. Overall, the results suggest that the scale should be applicable in both the U.S. and Singapore.

***Objective Measures***

Objective (true/false, multiple choice) measures were analyzed separately due to analytic considerations. Items can be found in SM Section 1. We employed Rasch modeling (Bond & Fox, 2007), a foundational model within item response theory (IRT), due to its theoretical robustness and practical advantages for measurement development. Importantly, Rasch modeling supports invariant measurement—item parameters are not sample-dependent, and person estimates are not test-dependent—enhancing the reliability and generalizability of the instrument across culturally and demographically diverse populations (DeMars, 2010).

We again first examined the item difficulty of the revised 20 *objective digital knowledge* items. After revising six from the original pool, difficulty ranged from $P$ = 0.49 to 0.98. Four items with persistently low difficulty were excluded. Additionally, one item with ambiguous wording was removed. A Rasch model was then fitted to the remaining 15 items; six were removed due to infit/outfit mean square (MSQ) values outside 0.5–1.5 (Linacre, 2002). Infit and outfit MSQ show how well responses fit the Rasch model. Infit is sensitive to unexpected responses near a person's ability level, while outfit is sensitive to unusual responses on very easy or very hard items (Bond and Fox, 2007). One additional item was dropped to balance true/false proportions. The final scale included eight items, with difficulty levels from $P = 0.49$ to 0.94, infit MSQ values of 0.80–1.03, and outfit MSQ values of 0.71–1.19. As shown in Figure 3 (upper), the items covered a good difficulty range relative to participant ability (IRT parameters = 1.51–1.82).

The difficulty of the nine items of the *objective information knowledge* ranged from $P = 0.43$ to $P = 0.94$ (M = .59, SD = .17). A Rasch model was fitted. One item with infit MSQ below 0.5 was removed. The final scale consists of 8 items, $P = 0.43$ to $P = 0.73$, infit MSQ values from 0.81 to 1.28, and outfit MSQ values from 0.72 to 1.38. As Figure 3 (bottom) shows, the items show a good range of item difficulties relative to participant abilities, with IRT item difficulty parameters ranging from -0.56 to 0.9.

***Convergent Validity***

To assess convergent validity, we conducted correlation analyses to examine whether the DMILS is associated with theoretically related constructs such as social media use and news consumption. Subjective DMILS was measured using the composite score of all 18 items, encompassing four subscales: DK, DS, IK, and IS (M = 3.54, SD = 0.67). Both the second-order confirmatory factor analysis, which supported a unidimensional structure, and the high internal consistency (Cronbach's $\alpha > .90$) justified the use of a single composite score. Objective DMILS was measured by summing the total score of 16 items, which included 8 true/false statements assessing objective digital knowledge and 8 multiple-choice items measuring objective information knowledge. Each correct response was scored as 1 and incorrect as 0, resulting in a total score ranging from 0 to 16 (M = 10.22, SD = 3.11). We included both composite and subscale scores in the analysis to evaluate the utility of using a unified measure versus its individual components.

Subjective DMILS and all four sub-dimensions were significantly positively correlated with social media use (See SM Section 8 for detailed statistics). In contrast, social media use was significantly negatively correlated with objective DMILS and its two subscales. This contrasting pattern may reflect a discrepancy between individuals' perceived and actual DMIL. Those who spend more time on social media may feel more confident navigating digital environments, leading to higher self-reported literacy, but this confidence does not necessarily translate to accurate knowledge or skills as measured objectively. News consumption was also significantly correlated with subjective DMILS, objective DMILS, and all six subscales. These small to moderate correlations, consistent with Cohen's (1992) guidelines, support the scale's convergent validity.

***Predictive Validity***

We conducted both OLS and Poisson regression analyses to assess the predictive validity of the DMILS in relation to discernment and information retrieval outcomes, including age and gender as covariates. Different modeling strategies were applied based on the distributional properties of the outcome variables. The first DV (discernment) ranged from 0 to 16 and, although count-based, showed a broader and approximately normal distribution. The second DV (information retrieval accuracy) ranged from 0 to 3 and was modeled using Poisson regression, appropriate for low-range count data with limited variance. Results are presented in Table 3.

In Model 1, both subjective and objective DMILS were significant predictors of discernment. Specifically, subjective DMILS showed a positive association ($\beta = 0.306$, $p = .012$), and objective DMILS was a stronger predictor ($\beta = 0.336$, $p < .001$). Model 2 decomposed the DMILS scores into their subdimensions. Among the subjective subscales, only IS significantly predicted discernment ($\beta = 0.443$, $p < .001$). Neither DK, DS, nor IK showed significant effects.

On the objective side, both objective DK ($\beta = 0.317$, $p < .001$) and objective IK ($\beta = 0.314$, $p < .001$) were strong, significant predictors of discernment.

Model 3 and Model 4 examined the predictive validity of the DMILS on participants' ability to retrieve accurate information. Model 3 showed that composite subjective and objective DMILS scores were both positively and significantly associated with the information retrieval accuracy. Specifically, a one-unit increase in subjective DMILS was associated with a 7.6% increase in the expected number of correct responses (log-IRR = 0.076, $p < .05$), while a one-unit increase in objective DMILS was associated with a 2.2% increase (log-IRR = 0.022, $p < .01$).

Model 4 entered the four subjective subcomponents and two objective sub-scores. None of the subjective subscales were significant predictors of retrieval accuracy. Similarly, neither objective DK nor objective IK significantly predicted it. These results suggest that while overall DMILS scores predict retrieval accuracy, the individual subcomponents may not independently explain significant variation when modeled together.

We also conducted separate analyses for each country to examine potential differences, including age, gender, ethnicity, and education as covariates (see Section 8 in SM); the findings were consistent with the main models. Overall, the findings provide evidence for the predictive validity of the DMILS across both discernment and information retrieval tasks.

**Discussion**

Study 2 provides strong evidence for the validity and cross-cultural applicability of the DMILS. The scale demonstrated good construct validity through CFA and Rasch modeling, with both subjective and objective components showing reliable and interpretable structures. Measurement invariance analyses confirmed that the scale functions equivalently across U.S. and Singapore samples, supporting its generalizability. The scale also showed expected associations with related constructs, such as news use and social media use, and significantly predicted both misinformation discernment and information retrieval accuracy.

## General Discussion

Although much work has been done to conceptualize and operationalize different aspects of media-related literacy (e.g., Cho et al., 2024; Koc & Barut, 2016), the fragmented nature of existing measures has hindered calls for a more comprehensive framework of digital media and information literacies (Jones-Jang et al., 2021). Operationalization is a crucial part of explicating concepts (McLeod & Pan, 2005), and for media literacy research to advance, measures must be easily adopted and adapted across different contexts to support both theoretical development and practical application, such as in media literacy education and intervention design.

Building on our multidimensional framework—encompassing domains (digital vs. information literacy), types of competencies (knowledge vs. skill), and measurement modes (objective vs. perceived)—we developed an 18-item subjective scale and a 16-item objective measure of Digital Media and Information Literacy, along with a short form (8 self-report and 8 objective items). Both were rigorously validated following established guidelines (Carpenter et al., 2018).

Results from EFA and CFA consistently supported this proposed four-factor structure, rather than collapsing into fewer dimensions, providing empirical evidence for the distinction between digital media literacy and information literacy, as well as between knowledge and skill within each domain. This finding carries important implications for how researchers conceptualize and study media-related literacies. It suggests that treating literacy as a monolithic construct, as is common in studies that rely on a single self-report measure, may obscure important variation in the specific competencies individuals possess. For instance, an individual may possess strong digital skills (e.g., navigating platforms, using search tools effectively) while lacking the information knowledge needed to evaluate the credibility of content encountered through those platforms.

This flexibility is especially important because different dimensions of DMILS may matter for different outcomes. The present results show that information skills are most predictive of misinformation discernment, an outcome that centrally involves evaluative and procedural competencies such as source checking and contextual assessment. However, other dimensions of DMILS may be more consequential for different outcomes. For example, prior research shows that higher levels of news literacy are associated with greater current events knowledge, stronger internal political efficacy, and more critical orientations toward political institutions (Ashley et al., 2017; Amazeen & Bucy, 2019; Vraga & Tully, 2021). Information literacy has also been linked to individual well-being through increased agency and reduced uncertainty (Lund & Wang, 2024). From this perspective, misinformation discernment represents one important, but non-exclusive, outcome domain in which DMILS may be consequential.

At the same time, second-order model analysis supports the unidimensional use of the scale, meaning that researchers may opt to generate a composite score by summing all item responses. This dual capability—subscale-level specificity alongside an overall composite—is a distinctive feature that most existing literacy measures lack. Importantly, this structure also enables the DMILS to function as a diagnostic tool for measuring and monitoring digital media and information literacy as an outcome itself.

By integrating both subjective and objective components within a single instrument, the DMILS addresses a critical limitation of prior measurement approaches by distinguishing perceived competence from demonstrated ability, allowing for a more precise assessment of how literacy functions in practice. Importantly, our findings show that objective media literacy measures are stronger predictors of misinformation discernment than subjective self-reports. This supports prior evidence (e.g., Jones-Jang et al., 2021) showing that only objective measures of information literacy were associated with improved detection of fake news, while self-reported measures of media, digital, and news literacy were not. It also enables researchers to quantify literacy "gaps" and examine how overconfidence in perceived literacy—similar to Dunning–Kruger effects—may shape sharing behavior and susceptibility to false content (Dunning, 2011).

The DMILS demonstrates cross-cultural applicability. Evidence of measurement invariance confirms that the scale functions equivalently across groups, and the consistency in convergent and predictive validity in both the U.S. and Singapore samples further supports its generalizability. This enables reliable monitoring and comparing individuals' digital media and information literacy across cultural contexts and over time. Given that misinformation and digital media challenges are global phenomena, the ability to deploy a common measure across national contexts is essential for building cumulative, comparative knowledge about how literacy operates in different information environments.

Beyond its contributions to measurement and theory, the DMILS has direct practical implications for the design and evaluation of media literacy interventions. Many existing media literacy programs adopt a general approach, aiming to improve overall critical thinking about media without targeting specific competencies. The multidimensional structure of the DMILS enables educators and practitioners to first assess which particular dimensions of literacy are weakest within a given population and then to design interventions that address those specific gaps. For example, if assessments using the DMILS reveal that a group scores high on digital skills but low on information knowledge, interventions can be tailored to focus on source evaluation and factual verification rather than on platform navigation skills they already possess. Furthermore, the inclusion of both subjective and objective components allows practitioners to identify populations that overestimate their own competencies—a pattern our results suggest is particularly common among heavy social media users. Recognizing this discrepancy is critical for intervention design, as individuals who believe they are already literate may be less receptive to educational programs and may require different pedagogical strategies, such as calibration exercises that reveal the gap between perceived and actual performance. In educational settings, instructors could administer the DMILS as a pre- and post-assessment to evaluate program effectiveness across specific competency areas, enabling evidence-based refinement of curricula over time.

While designed to be broadly applicable, media literacy remains influenced by sociocultural and economic contexts. Researchers should carefully consider the populations they are studying and may benefit from complementing the DMILS with qualitative methods to capture aspects that are difficult to assess quantitatively. Additionally, the current samples exhibited relatively high levels of media literacy, as reflected in both subjective and objective assessments, likely due to the use of online crowdsourcing platforms for recruitment. Future research could employ alternative sampling strategies. Analysis of the inter-person item map and item characteristic curve (ICC) plots suggests that the objective measure may have been too easy for many participants, potentially limiting its ability to discriminate among individuals with higher levels of digital literacy. To address this limitation, future iterations of the scale could benefit from incorporating more challenging items, such as evaluation of complex scenarios (e.g., detecting more nuanced phishing attempts).

In conclusion, this study developed and validated a comprehensive DMILS that integrates both subjective and objective dimensions of literacy across digital and informational domains. Through rigorous psychometric testing, DMILS demonstrated strong reliability, theoretical coherence, and practical utility. Its cross-cultural robustness further underscores its potential as a versatile tool for assessing media literacy across diverse populations. By capturing both perceived competence and actual ability, the DMILS offers valuable insights into literacy gaps and supports future research and intervention design aimed at enhancing individuals' ability to navigate today's complex information environment.

## Appendix I

### Subjective DMILS – Short Version (8 items)

**1. Digital Knowledge (DK)**

*Instruction: How familiar are you with the following computer and Internet-related items? (1= No Understanding, 2 = Slight Understanding, 3 = Moderate Understanding, 4 = Substantial Understanding, 5 = Full Understanding)*

1. Phishing
2. Algorithm

**2. Digital Skill (DS)**

*Instruction: Please select how much you agree or disagree with the following statements. (Strongly disagree to Strongly agree)*

1. I know how to solve my own technical problems on digital devices and platforms.
2. Using any technological device comes easy to me.

**3. Information Knowledge (IK)**

*Instruction: Please select how much you agree or disagree with the following statements. (Strongly disagree to Strongly agree)*

1. I recognize the norms that underlie journalists' work.

2. I understand the routines in which journalists engage in reporting and content creation.

**4. Information Skill (IS)**

*Instruction: Please select how much you agree or disagree with the following statements. (Strongly disagree to Strongly agree)*

1. I look for more information before I believe something I see in the news.
2. I often consider whether a message in news is accurate.

**Objective DMILS – Short Version (8 items)**

**1. Objective Digital Knowledge**

1. Reverse image search is a feature that flips images horizontally. (F)

2. Chatbots always provide accurate and reliable information in response to user queries.

(F)

3. Machine learning is a branch of artificial intelligence that involves training computer systems to learn from data and improve over time without being explicitly programmed (T).

4. Spyware is software used by intelligence agencies to gather classified information. (F)

**2. Objective Information Knowledge**

1. The most reliable, verified, concise and comprehensive description of an unknown specialized concept can be found in _____

a. daily newspaper
b. bilingual dictionary
**c. lexicon or encyclopedia**
d. research article

2. It has been scientifically established that cholesterol is present in animal organisms but not in plants. How would you best describe a TV commercial which claims that the sunflower oil manufactured by a particular producer contains no cholesterol?
a. This is a valuable benefit, and it will encourage me to buy this brand of oil.
**b. This is manipulative and misleading information, as plant oils do not contain cholesterol.**
c. This information has medical significance, and I am therefore willing to pay more for this oil.
d. This is interesting information on the unique composition of this oil.

3. Which of the following is the LEAST useful when evaluating the credibility of a piece of information?

a. The authority of the article’s author
b. How up-to-date the information is
c. The point of view of the author
**d. The visual appeal of the design elements**

4. Which of the following is typically responsible for writing a press release
a. A reporter for a news organization
**b. A spokesperson for an organization**
c. A lawyer for a news aggregator
d. A producer for a news organization

Table 1. Multidimensional framework of digital media and information literacy (DMIL)

| | | **Digital Media Literacy** | **News/Information Literacy** |
|---|---|---|---|
| **Subjective-DMIL** | Knowledge | *Digital Knowledge* | *Information Knowledge* |
| | | Self-reported familiarity with tech-related terminology (Hargittai & Hsieh, 2012). | Self-reported understanding of how news and information are produced, structured, and shared (Vraga et al., 2021). |
| | Skill | *Digital Skill* | *Information Skill* |
| | | Self-reported ability to operate, navigate, and create content using digital tools and technologies (Guess & Munger, 2023; Koc & Barut, 2016; Ng, 2012) | Self-reported ability to critically evaluate news and information, verify content, and engage with media through informed judgment or social interaction (Koc & Barut, 2016) |
| **Objective-DMIL** | Knowledge | *Objective Digital Knowledge* | *Objective Information Knowledge* |
| | | Factual knowledge of technology and digital tools | Factual knowledge of news production processes (Anspach & Carlson, 2024; Boh Podgornik et al., 2016) |

Table 2. Results of exploratory factor analysis of the Subjective DMILS.

| Factor/Item | Factor loading | Variance explained | Cronbach's alphas |
|---|---|---|---|
| *Digital Knowledge (DK)* | | 15.94% | 0.82 |
| DK1: How familiar are you with PDF? | 0.6 | | |
| DK2: How familiar are you with Spyware? | 0.82 | | |
| DK3: How familiar are you with Phishing? | 0.76 | | |
| DK4: How familiar are you with Hashtag? | 0.58 | | |
| DK5: How familiar are you with Chatbot? | 0.73 | | |
| DK6: How familiar are you with Algorithm? | 0.62 | | |
| *Digital Skill (DS)* | | 13.61% | 0.85 |
| DS1: I know how to solve my own technical problems on digital devices and platforms. | 0.76 | | |
| DS2: I can learn new technologies easily. | 0.88 | | |
| DS3: I have the technical skills I need to use information and communication technologies (ICT) for learning and to create artifacts (e.g., presentations, digital stories, wikis, blogs). | 0.65 | | |
| DS4: Using any technological device comes easy to me. | 0.79 | | |
| *Information Knowledge (IK)* | | 13.62% | 0.85 |
| IK1: I understand the legal protections and constraints of journalism. | 0.8 | | |
| IK2: I recognize the norms that underlie journalists' work. | 0.85 | | |
| IK3: I understand the routines in which journalists engage in reporting and content creation. | 0.78 | | |
| IK4: I understand how news is made. | 0.66 | | |
| *Information Skill (IS)* | | 12.33% | 0.83 |
| IS1: I compare new information I see in news with other information I have seen before accepting it as believable. | 0.7 | | |
| IS2: I look for more information before I believe something I see in the news. | 0.78 | | |
| IS3: It is important to think twice about what news messages say. | 0.68 | | |
| IS4: I often consider whether a message in news is accurate. | 0.8 | | |

Table 3. Predictive validity of DMILS

| | Discernment *OLS* | | Information Retrieval *Poisson* | |
|---|---|---|---|---|
| | (1) | (2) | (3) | (4) |
| Subjective DMILS | 0.306$^{*}$ | | 0.076$^{*}$ | |
| | (0.122) | | (0.036) | |
| Objective DMILS | 0.336$^{***}$ | | 0.022$^{**}$ | |
| | (0.025) | | (0.007) | |
| Digital knowledge | | 0.129 | | 0.015 |
| | | (0.094) | | (0.028) |
| Digital skill | | -0.021 | | 0.061 |
| | | (0.119) | | (0.036) |
| Info knowledge | | -0.127 | | -0.012 |
| | | (0.109) | | (0.032) |
| Info skill | | 0.443$^{***}$ | | 0.017 |
| | | (0.133) | | (0.040) |
| Objective Digital | | 0.317$^{***}$ | | 0.028 |
| | | (0.053) | | (0.016) |
| Objective Info | | 0.314$^{***}$ | | 0.015 |
| | | (0.046) | | (0.014) |
| Constant | 5.223$^{***}$ | 4.849$^{***}$ | 0.335$^{*}$ | 0.275 |
| | (0.507) | (0.621) | (0.151) | (0.186) |
| Observations | 998 | 998 | 998 | 998 |
| $R^2$ | 0.199 | 0.208 | | |
| Adjusted $R^2$ | 0.195 | 0.200 | | |
| Log Likelihood | | | -1,567.559 | -1,566.390 |
| Akaike Inf. Crit. | | | 3,147.118 | 3,152.779 |
| Residual Std. Error | 2.282 (df = 992) | 2.274 (df = 988) | | |
| F Statistic | 49.317$^{***}$ (df = 5; 992) | 28.756$^{***}$ (df = 9; 988) | | |

Note: $^{*}p < 0.05$; $^{**}p < 0.01$; $^{***}p < 0.001$. Models 1 and 3 include the overall subjective and objective DMILS scores, while Models 2 and 4 include the subcomponents of the subjective and objective DMILS scores. In Models 1 and 2, coefficients represent the change in the discernment score. In Model 3 and 4, coefficients represent log incidence rate ratios. Standard errors are shown in parentheses. In Models 1 and 2, one participant was excluded due to missing data on an image-based discernment item, resulting in an analytic sample of 998. In Model 3 and 4, one participant was excluded due to missing data on an information retrieval question, resulting in an analytic sample of 998. Gender and age were included as covariates in the two models. Variance inflation factor (VIF) diagnostics indicated no evidence of problematic multicollinearity (all VIFs < 2).

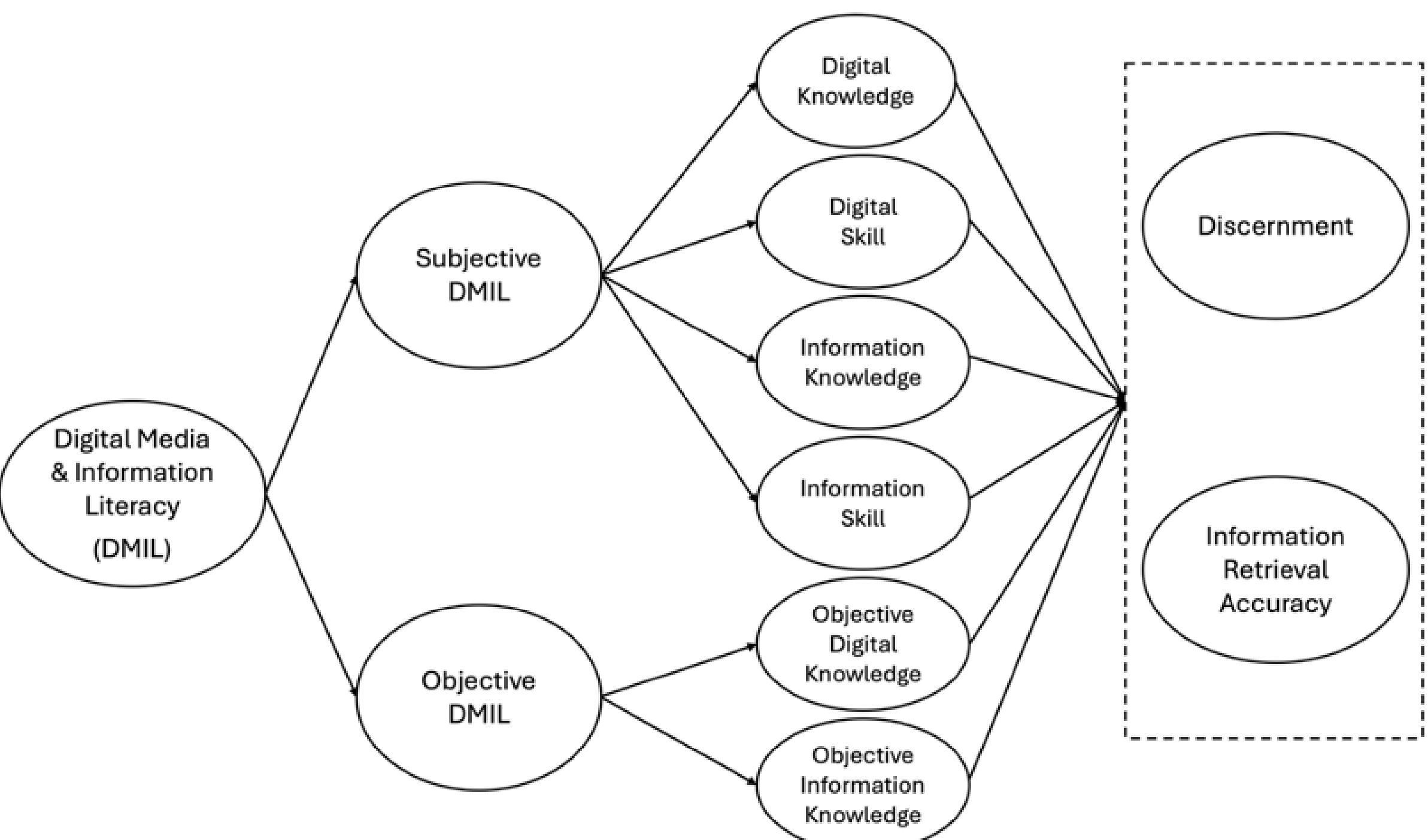

**Figure 1.** Measurement model

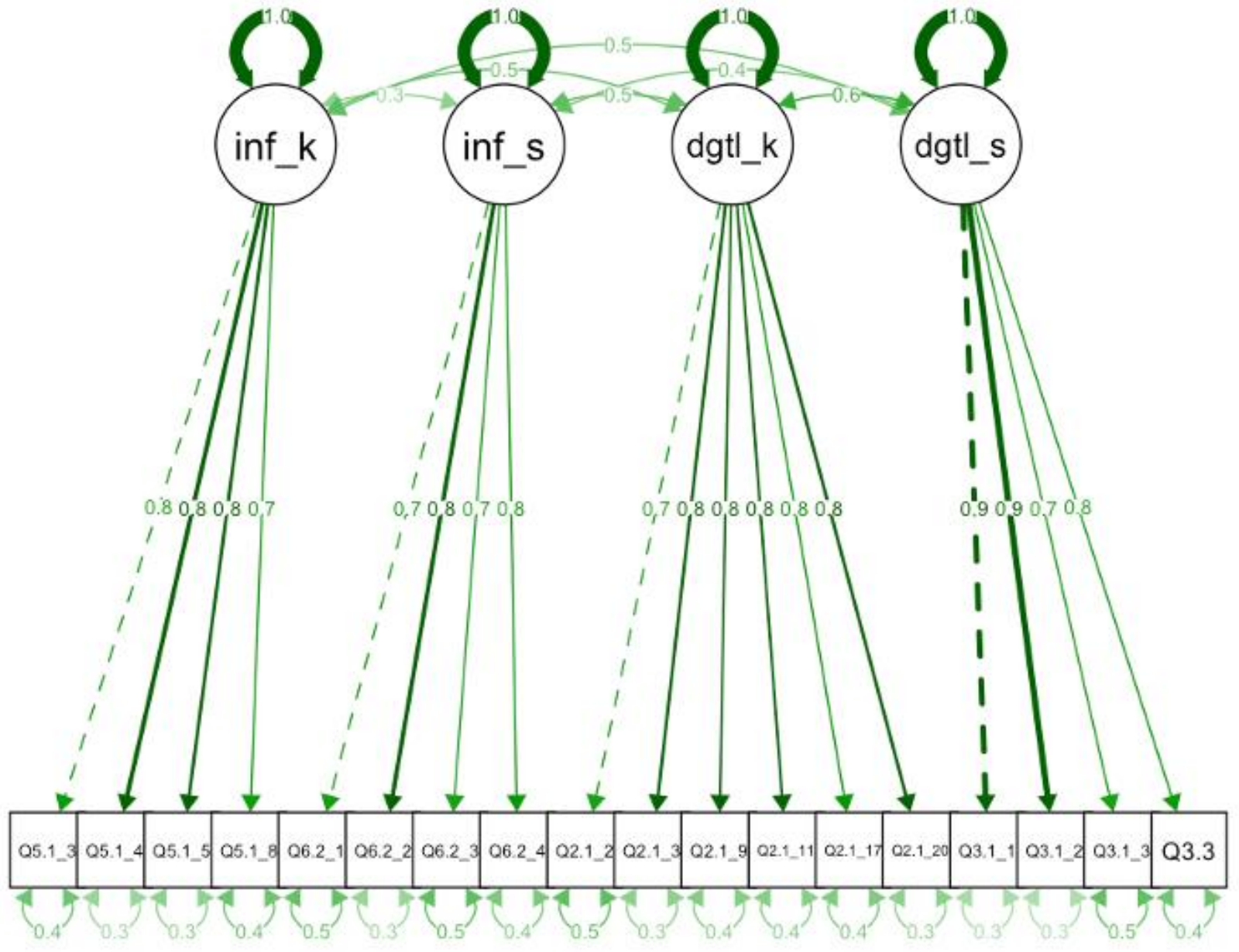

**Figure 2.** Results of confirmatory factor analysis of the four-factor model of Subjective DMILS

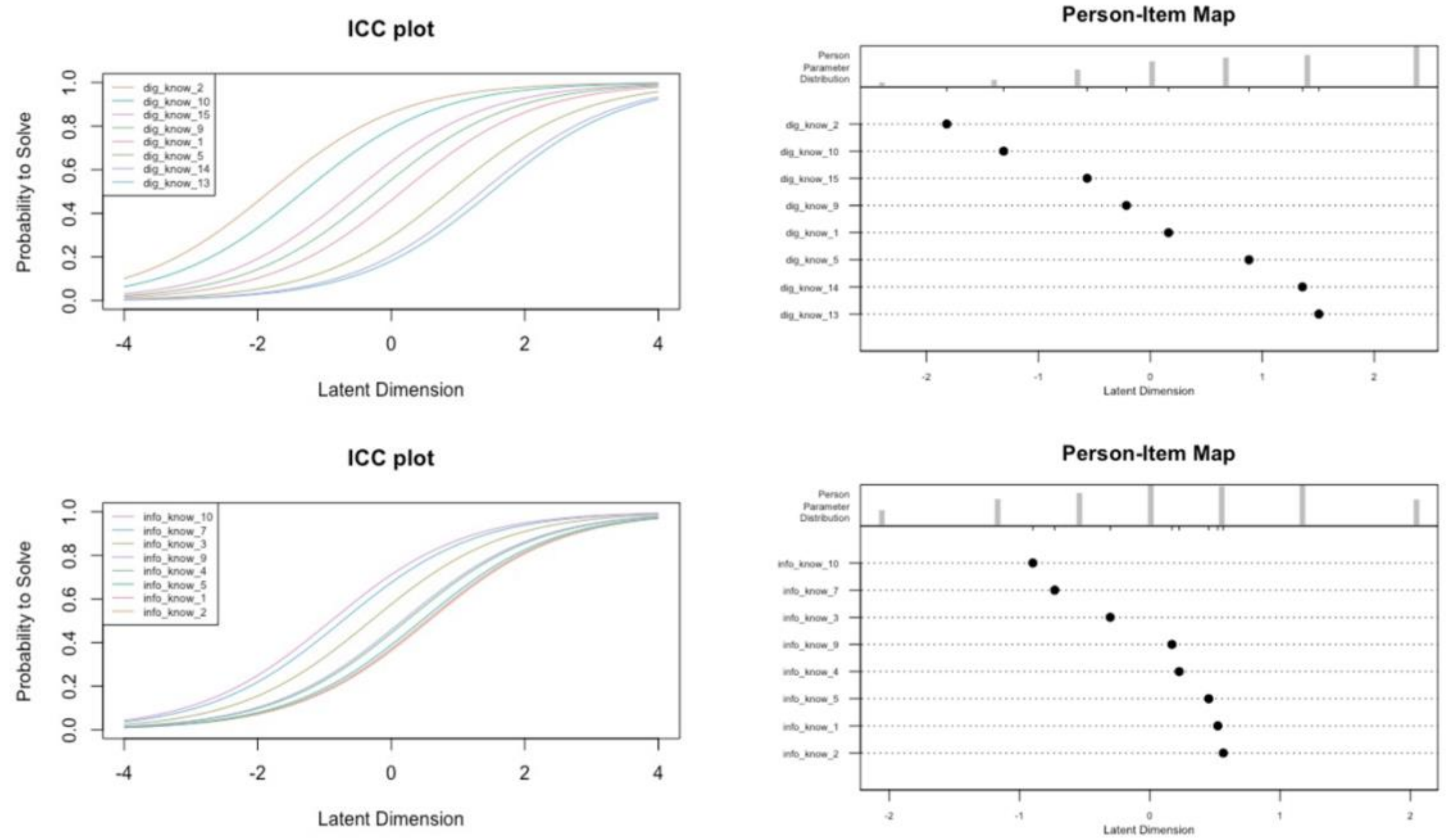

**Figure 3.** Item characteristic curves (ICCs) and person–item maps for digital knowledge items (upper) and information knowledge items (lower).